\documentclass[a4paper, oneside, twocolumn, notitlepage, 10pt]{extarticle_ecoc}
\usepackage{ecoc}
\usepackage[absolute]{textpos}
\usepackage{caption}
\usepackage{subcaption}

\begin{document}
\selectlanguage{english}    

\title{Metro Access Network with Convergence of Coherent and Analog RoF Data Services}%

\author{
    Amol Delmade\textsuperscript{(1*)}, Frank Slyne\textsuperscript{(2)}, Colm Browning\textsuperscript{(1,3)}, Daniel Kilper\textsuperscript{(2)}
    Liam Barry\textsuperscript{(1)} and Marco Ruffini\textsuperscript{(2)}
}

\begin{textblock}{14}(1,1)
\noindent This paper is a preprint of a paper accepted to ECOC 2023 and is subject to Institution of Engineering and Technology Copyright. The copy of record is available at IET Digital Library.
\end{textblock}

\maketitle                  

\begin{strip}
 \begin{author_descr}

   \textsuperscript{(1)} School of Electronic Engineering, Dublin City University, Ireland \textcolor{blue}{\uline{*amol.delmade2@mail.dcu.ie}}

   \textsuperscript{(2)} CONNECT Center, School of  Computer Science and Statistics, Trinity College Dublin, Ireland
   
   \textsuperscript{(3)} mBryonics Ltd., Unit 13 Fiontarlann Teo, Westside Enterprise Park, Galway, H91 XK22, Ireland

 \end{author_descr}
\end{strip}

\setstretch{1.1}


\begin{strip}
  \begin{ecoc_abstract}
Efficient use of spectral resources will be an important aspect for converged access network deployment. This work analyzes the performance of variable bandwidth Analog Radio-over-Fiber signals transmitted in the unfilled spectral spaces of telecom-grade ROADM channels dedicated for coherent signals transmission over the OpenIreland testbed.
  \end{ecoc_abstract}
\end{strip}


\section{Introduction}

The convergence of different data services and efficient use of spectral resources over metro and access networks will be an important aspect of the next-generation communication systems. Coherent optical links have extended their use from long-haul to metro networks \cite{Ref1}, while spectrally inefficient digital radio-over-fiber (DRoF) links are currently used for fronthauling of the wireless signals to remote antenna sites \cite{Ref2}. The deployment of spectrally efficient fronthaul networks and sharing of resources will be paramount to meet the increasing densification and capacity demands with 5G/6G technologies \cite{Ref3,Ref3_1}. Analog radio-over-fiber (ARoF) transmission has shown promising results for efficient fronthaul implementation as it retains the spectral efficiency (SE) of wireless signals \cite{Ref4,Ref5,Ref6}. 
The high SE of ARoF is exploited here for convergence with coherent signals.

Typically a coherent signal does not occupy the complete bandwidth (BW) allocated by a reconfigurable optical add-drop multiplexer (ROADM), as they exhibit a 6.25 GHz or higher spectral granularity. In this work, variable BW ARoF signals were transmitted in the unfilled spectral spaces of ROADM channels carrying coherent traffic. We combined the open networking multi-vendor components from the OpenIreland Testbed at Trinity College Dublin with hybrid RF/optical research facilities at Dublin City University to examine the interaction between coherent and ARoF signals. In the previous work \cite{Ref8}, we demonstrated the convergence of a 2.93 Gb/s orthogonal frequency division multiplexing (OFDM) 488 MHz BW ARoF and 100 Gb/s DP-QPSK coherent signals over 49 km distance with signals separated in sub-channels within the primary 50 GHz ROADM channel. In this work, we eliminated the sub-channelization to allow closer cohabitation of signals. In order to further increase the spectral usage, we added an additional ARoF signal on the lower frequency empty spectrum of the ROADM channel while increasing the OFDM signal's BW to 1.6 GHz. The successful transmission of high BW signals, to remote antenna sites, will be important as the wireless network moves toward 6G \cite{Ref3}. In this work, we demonstrated the transmission of 2$\times$9.2 Gb/s OFDM A-RoF signals in convergence with 100 and 400 Gb/s coherent signals within 50 and 100 GHz ROADM channels, respectively. To the best of our knowledge, this is the first successful convergence demonstration of high BW ARoF and 100/400 Gb/s coherent signals with telecom-grade components.
Convergence of pulse amplitude modulation (PAM) and ARoF signals has been demonstrated in \cite{Ref10, Ref11}, while DRoF-ARoF convergence through a multi-$\lambda$ network is demonstrated in \cite{Ref12}.

\vspace{-7 pt}
\section{Experiment Details}
\begin{figure*}
\centering
  \includegraphics[width=14cm,height=8cm]{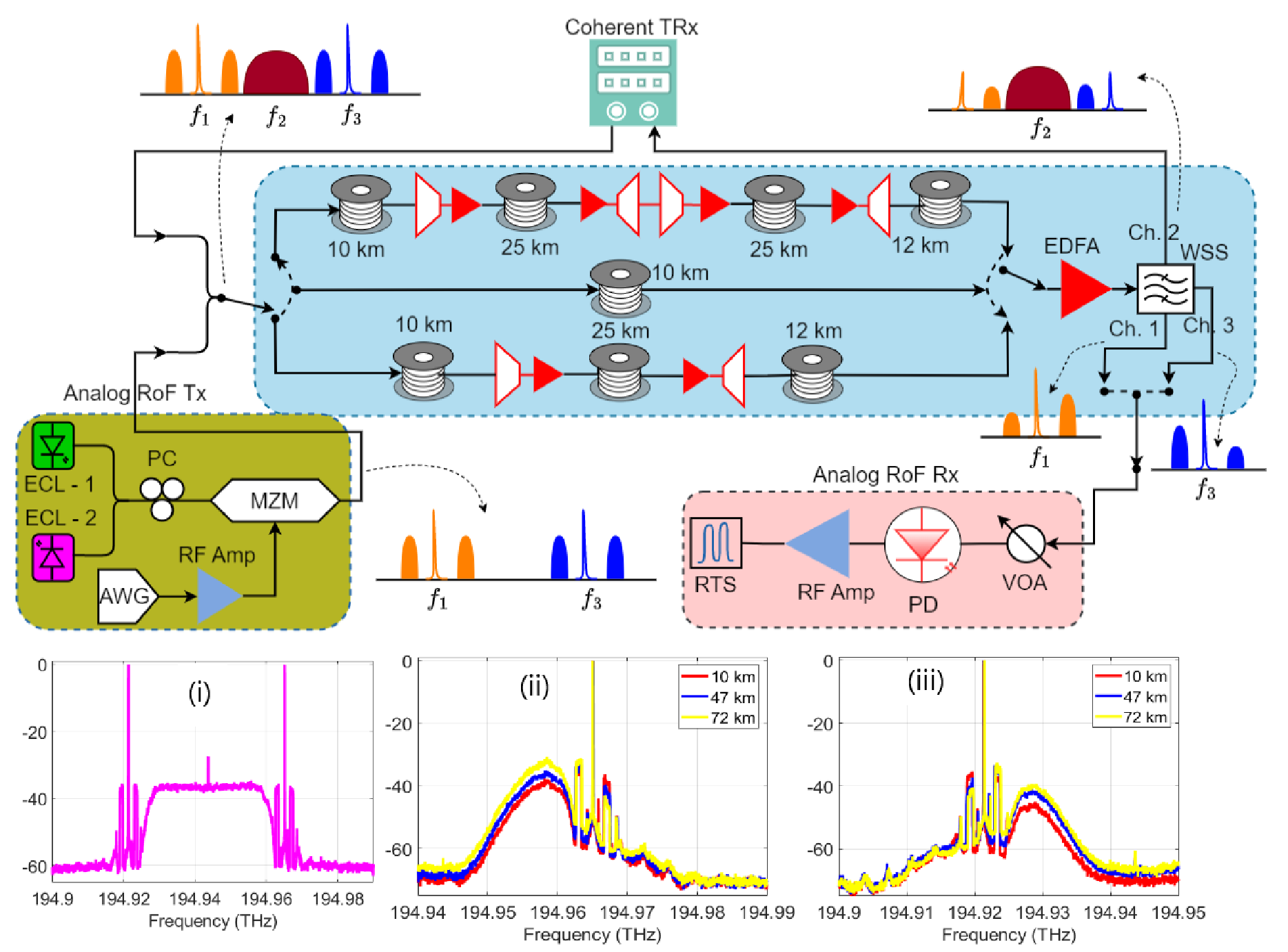}
  \caption{Experimental testbed for the convergence of coherent and ARoF data with insets showing the optical spectrum of the (i) combined signal; WSS filtered (ii) low and (iii) high frequency ARoF signals for 100Gb/s coherent and 800 MHz BW ARoF case.}
  \label{fig:architecture}
\end{figure*}

The experimental testbed showing the generation, combination and detection of two A-RoF OFDM signals in close proximity to a coherent signal is shown in Fig.\ref{fig:architecture}. The use of open networking telecom-grade components from the OpenIreland testbed allowed a programmable configuration of the physical light path topology and power monitoring throughout the link. 

A commercial Teraflex FSP3000C transponder from ADVA was employed for the generation and detection of 100 and 400 Gb/s coherent signals. The 100 Gb/s (31.50 Gbaud DP-QPSK) signal occupies 37.64 GHz BW within a 50 GHz ROADM channel, while the 400 Gb/s signal (69 Gbaud DP-16-QAM) occupied an 82.46 GHz BW within a 100 GHz ROADM channel. The ROADM BW values were specified by the transponder vendor for regular metro network operation. Two ARoF signals were added, on either side of the coherent data (spectrum shown in Fig.\ref{fig:architecture}(i)), using an external coupler to fill the empty spectrum.

The ARoF signals were generated using two tunable external cavity lasers (ECLs) and an external Mach-Zehnder modulator (MZM), modulated by intermediate frequency (IF) OFDM data, as shown in Fig.\ref{fig:architecture}. The ECLs were tuned such that the generated ARoF signals occupy the empty spectum within the assigned ROADM BW. The same IF OFDM data was modulated on both lasers due to the availability of a single MZM. Offline-generated IF OFDM signals were converted to the electrical domain using an arbitrary waveform generator (AWG). 
The 200, 400, 800 MHz and 1.6 GHz BW OFDM signals were generated by changing the number of data subcarriers (SC). A 64-QAM data modulation on SCs resulted in the minimum and maximum data rates of 1.2 to 9.6 Gb/s for 200 MHz and 1.6 GHz BW signals, respectively. The generated dual sideband ARoF signals, with an IF of 2 GHz and maximum signal BW of 1.6 GHz, occupied $\sim$6 GHz of ROADM BW on either side of the coherent data.

The combined signal, consisting of two A-RoF and one coherent waveform (spectrum shown in Fig.\ref{fig:architecture}(i) for the 100 Gb/s case), was transmitted through three different light path topologies, with different distances and number of ROADMs. Topology A consists of a 10 km fiber loop with no ROADMs, while topologies B and C extend to 47 and 72 km with one and two ROADMs, respectively. The 10 and 12 km fibres added at the beginning and end of the path for topologies B and C, account for the fact that the antenna site and data processing site, respectively, might not be colocated with the central office hosting the ROADMs. In each topology, the received signal was amplified using an Erbium-doped fiber amplifier (EDFA) before being routed to a wavelength selective switch (WSS), which employs a programmable filter profile to differentiate between the signals. The ARoF and coherent signals were then routed to the respective receivers as shown in Fig.\ref{fig:architecture}. The optical spectra of the WSS-filtered low and high-frequency 800 MHz BW ARoF signals are shown in Fig.\ref{fig:architecture}(ii) and (iii), respectively, for 100 Gb/s coherent signal convergence case. 

The ARoF receiver consists of a photodetector (PD), an RF amplifier and a real-time oscilloscope (RTS). Offline processing was carried out on the captured IF OFDM signals to analyze the error vector magnitude (EVM) performance. 
\vspace{-7 pt}
\section{Results and Discussion}

\begin{figure*}[t]
\centering
  \subfloat[]
           {\includegraphics[width=0.45\linewidth,height=4.65cm]{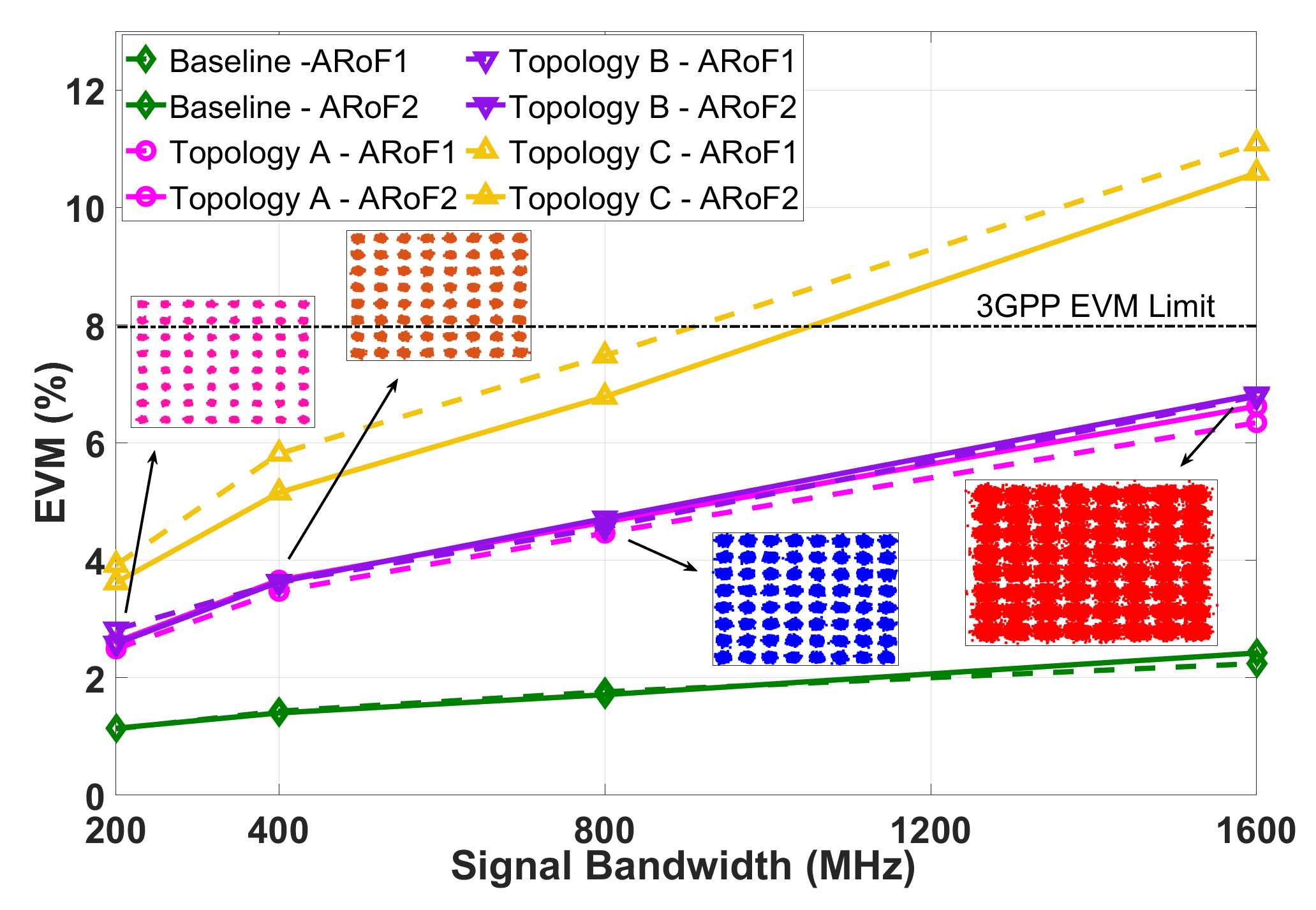} }
  \subfloat[]
            {\includegraphics[width=0.45\linewidth,height=4.65cm]{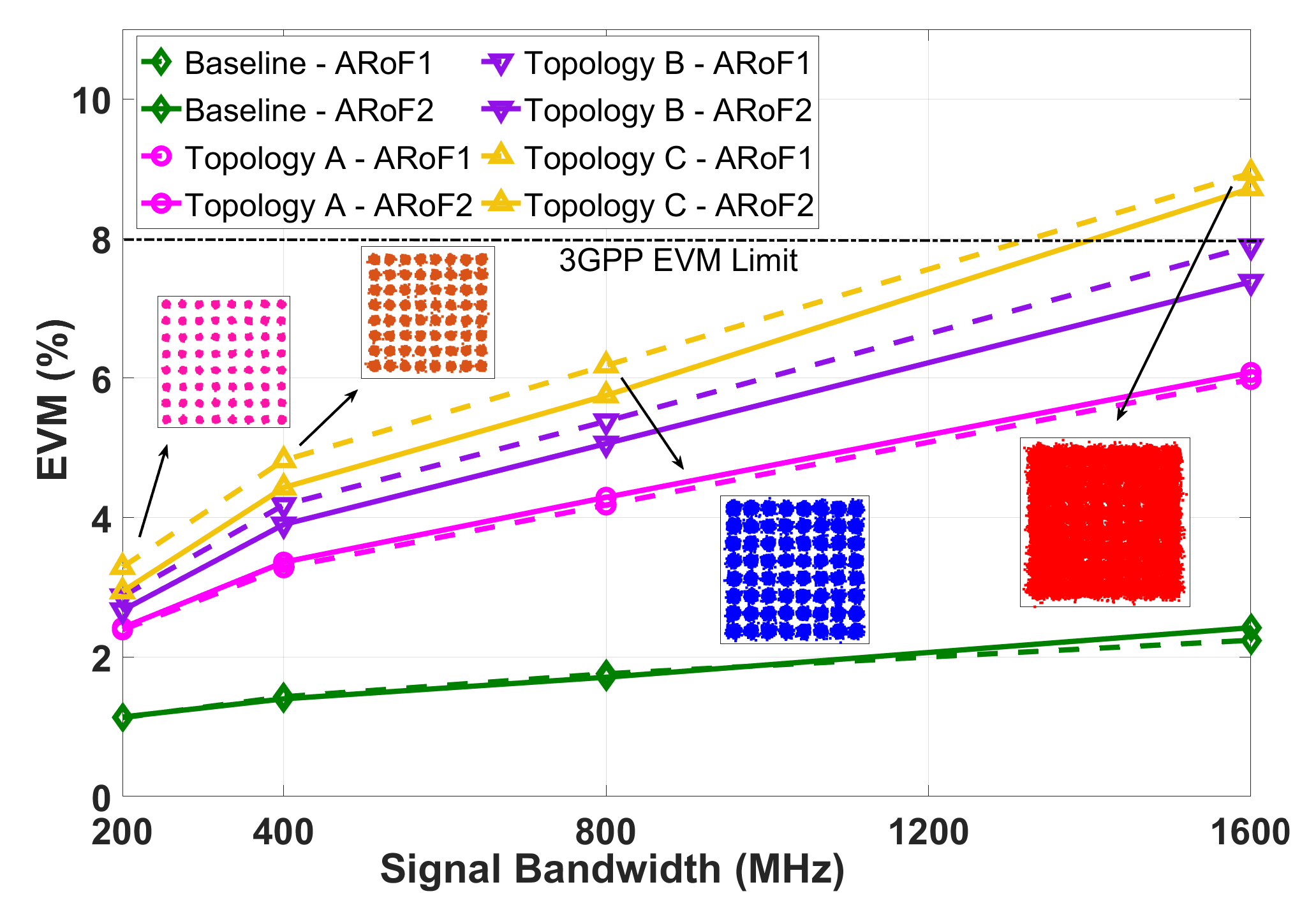}}
\vspace{1em}
\caption{EVM vs BW performance for A-RoF OFDM signals in convergence with (i) 100Gb/s and (ii) 400Gb/s coherent signals.} 
\label{fig:results_1}
\end{figure*}

The EVM performances of variable BW ARoF signals (low and high frequency; ARoF1 and ARoF2) for three transmission topologies are shown in Fig.\ref{fig:results_1}(i) and (ii), in convergence with 100 and 400 Gb/s coherent signals, respectively. An increase in the signal BW from 200 MHz to 1.6 GHz degraded the performance even for the baseline back-to-back (B2B) transmission case, without the coherent signals and WSS, as seen from the green curves. This degradation is reflected in the constellations of ARoF signals shown as insets in Fig.\ref{fig:results_1}(i) and (ii) for different transmission cases.  

The addition of coherent signals and WSS filtering, for topology A, resulted in performance degradation as shown by different EVM curves for topology A (pink) and baseline (green) in Fig.\ref{fig:results_1}(i) and (ii). The 200 MHz BW OFDM signals EVM degraded from the baseline value of 1.3$\%$ to 2.49$\%$ for topology A in convergence with 100 Gb/s coherent data, while a significant $\sim$4$\%$ degradation was observed for the 1.6 GHz BW signal. The WSS filtering reduces the power of one data sideband, as seen from the spectra in Fig.\ref{fig:architecture}(ii) and (iii), resulting in performance degradation. Increasing the transmission distance with B and C topologies further degraded the EVM performance (violet and yellow curves) as a consequence of the reduced ARoF signals power by ROADMs non-ideal filtering. The performance of all BW OFDM signals is under the 8$\%$ 3rd generation partnership project (3GPP) EVM limit of 64-QAM \cite{Ref9} for B topology with both converged cases - indicating the successful transmission of 2$\times$9.2 Gb/s A-RoF OFDM signals up to 47 km. 

Different EVM values of ARoF signals, as seen from the yellow curves in Fig.\ref{fig:results_1}(i) and (ii), show the impact of convergence with different data rate coherent signals. A 400 Gb/s coherent data standard ROADM channel allows converged ARoF signals with BWs expandable up to almost 1.6 GHz. This bandwidth expansion is hindered to some degree through convergence with a 100 Gb/s coherent signal ($\sim$11$\%$ EVM for 1.6 GHz BW signal for topology C). This disparity in performance is caused by the different percentages of ROADM BW being allocated to the ARoF signals. Both ARoF signals are required to fit in a 12.36 GHz spectrum for convergence with the 100 Gb/s coherent signal. This free spectrum allocation increases to 17.54 GHz for the 400 Gb/s convergence case owing to the fact that a 100GHz channel spacing is specified for this coherent service. It can be inferred that the ARoF signal BW to ROADM channel BW ratio is the key performance deciding factor in such converged scenarios. 



The Q-factor values of the coherent signals were found to be slightly degraded with the introduction of WSS filtering in topology A, however its value did not change as we move to topologies B and C. The WSS filtering was applied to the coherent data to separate it from ARoF signals - slightly affecting its performance compared to the baseline case. The coherent signals do not experience any degradation on passing through the RODAMs, in topologies B and C, as they occupy a portion of the RODAM channels central BW with a flat passband - keeping the performance constant. The Q-factor values of 16.5 and 9.1 were measured for all three topologies for 100 and 400 Gbps coherent data, respectively, while the baseline values of 17.3 and 10.3 were observed.

\vspace{-7 pt}
\section{Conclusions}
The convergence of different data services and efficient use of spectral resources is paramount to fulfill the increasing data demand over the metro access network. This work demonstrates the use of telecom-grade equipments for the convergence of analog RoF and digital coherent data signals within ROADM channels. 
Various transmission scenarios with a combination of 100 and 400 Gb/s coherent data; 200 MHz to 1.6 GHz BW OFDM ARoF data signals and 10, 47 and 72 km transmission distances were implemented. Successful transmission of 2$\times$9.2 Gb/s and 2$\times$4.8 Gb/s A-RoF OFDM signals in convergence with 100 and 400 Gb/s coherent data is achieved up to 47 km and 72 km, respectively. The flexibility offered by the demonstrated system for the convergence of different data services and applications pave the way for its deployment in the metro network.

\vspace{-7 pt}
\section{Acknowledgements}
This work is funded by SFI 18/RI/5721, 13/RC/2077\_p2, 13/RC/2076\_p2, 18/SIRG/5579 and DTIF  EI DT2019 0014B grants.


\printbibliography


\end{document}